\documentclass[conference]{IEEEtran}
\IEEEoverridecommandlockouts
\usepackage{cite}
\usepackage{amsmath,amssymb,amsfonts}
\usepackage{algorithmic}
\usepackage{graphicx}
\usepackage{textcomp}
\usepackage{xcolor}
\usepackage{multirow}
\begin{document}

\title{Exploring Emerging Trends in 5G Malicious Traffic Analysis and Incremental Learning Intrusion Detection Strategies\\}

\author{\IEEEauthorblockN{Zihao Wang, Kar Wai Fok, Vrizlynn L. L. Thing}
\IEEEauthorblockA{\textit{Cybersecurity Strategic Technology Centre} \\
\textit{Singapore Technologies Engineering}}
}
\maketitle
 
\begin{abstract}
The popularity of 5G networks poses a huge challenge for malicious traffic detection technology. The reason for this is that as the use of 5G technology increases, so does the risk of malicious traffic activity on 5G networks. Malicious traffic activity in 5G networks not only has the potential to disrupt communication services, but also to compromise sensitive data. This can have serious consequences for individuals and organizations. In this paper, we first provide an in-depth study of 5G technology and 5G security. Next we analyze and discuss the latest malicious traffic detection under AI and their applicability to 5G networks, and compare the various traffic detection aspects addressed by SOTA. The SOTA in 5G traffic detection is also analyzed. Next, we propose seven criteria for traffic monitoring datasets to confirm their suitability for future traffic detection studies. Finally, we present three major issues that need to be addressed for traffic detection in 5G environment. The concept of incremental learning techniques is proposed and applied in the experiments, and the experimental results prove to be able to solve the three problems to some extent. 

\end{abstract}


\section{\bf Introduction}
With the rapid development of communication technologies, 5G has emerged as a key technology trend in recent years. It offers very high data rates and higher coverage with significantly improved quality of service (QoS), and very low latency ~\cite{agiwal2016next}, therefore attracting much attention. 5G wireless networks provide ultra-reliable broadband access to users through dense base station deployments, which include traditional cellular handheld devices, as well as a plethora of new devices related to machine-to-machine communication (M2M), Internet of Things (IoT), and cyber-physical systems (CPS) ~\cite{kutscher2016s} ~\cite{mitra20155g}. This has enabled many new applications that rely heavily on high network speeds and low latency, such as self-driving vehicles~\cite{hakak2023autonomous}, remote diagnosis and surgery ~\cite{cabanillas20235g}, and virtual reality (VR) ~\cite{zhai20215g}, which have greatly improved people's lives. In fact, According to the Thales 5G Progress Report, more than 100 5G networks were launched worldwide in 2020, and the number of 5G connections is expected to reach 1 billion by the end of 2023 ~\cite{1}. By 2030, more than 5 billion 5G connections are expected globally, driving nearly \$1 trillion in GDP growth ~\cite{2}. 5G continues to scale faster than any previous generation of mobile devices~\cite{34}.

The richness of 5G technology in terms of application areas also means that it is not a simple upgrade from 4G, but rather an integration of many of the latest disruptive technologies to meet the growing demand for the internet of everything and emerging smart and digital services in the future ~\cite{inproceedings}. In general, 5G integrates communication technology, information technology and operation Technology ~\cite{kao2022opportunities}. Therefore, in addition to the communication network itself, the network cloudification process involves the integration of hardware and software cloud vendors into the supply chains of network equipment. This has led to the entry of major hyperscalers, such as Microsoft and Google into the telecommunications industry to provide core and edge cloud solutions. New terminals that were originally limited by past generations of mobile networks, such as autonomous driving, VR/Augmented Reality (AR) technologies, more complex environmental monitoring, and robotic services will have a great opportunity to be realized with the support of 5G technology ~\cite{maulani2023development}. However, the increasing popularity of 5G technology also means that it becomes more susceptible to cyber threats which is a growing concern. The high data rates and low latency of 5G technology make it an attractive target for malicious actors seeking to conduct activities such as distributed denial of service (DDoS) attacks, botnet propagation, and data theft ~\cite{ahmad2019security}. These attacks not only have the potential to disrupt communication services, but also compromise sensitive data, resulting in serious consequences for both individuals and organizations.

5G adopts new technological concepts or solutions such as cloud computing ~\cite{rost2014cloud}, Software Defined Networking (SDN) ~\cite{hu2014survey}, and Network Function Virtualization (NFV) ~\cite{han2015network} to meet the needs of increasingly diverse applications and connected devices. However, security challenges associated with each of these technologies have been identified in 5G networks. SDNs that centralize network control logic in SDN controllers could be a prime target for attackers to launch denial-of-service (DoS) or resource exhaustion attacks, leading to network paralysis. Similarly, management programs in NFV are also vulnerable to such attacks. The cost of recovering from attacks caused by the lack of proactive 5G network security mechanisms will be far greater than the cost of proactive network defense up front. The 2017 NotPetya attack which caused \$10 billion in enterprise losses, including over \$1 billion in combined losses for Merck, Maersk, and FedEx ~\cite{3} is a strong testament to this. While 5G networks did not exist at that time, this attack illustrates and foreshadows the potential for even more irreparable damage in future 5G network with the Internet of Everything. In this context, analyzing complex network traffic to identify malicious traffic activity become even more critical.

In traditional network environments, network traffic detection can be divided into two main categories. The first is the traditional payload-based technique, such as deep packet inspection (DPI) and port-based identification. The other is based on artificial intelligence techniques such as traditional machine learning based or deep learning based traffic detection and classification. For the first category, with the increase in network bandwidth and the influx of large amounts of data into the Internet, the network traffic identification feature rules that need to be stored in the rule base library for traffic identification are rapidly expanding. The processing and storage costs of the system are also increasing. More importantly, complete network payload analysis is not only computationally expensive, but may also involve user privacy disputes and data security leaks ~\cite{wang2022machine}. In particular, decryption of the encrypted network traffic so as to inspect the payload information actually becomes a malicious act. According to google transparency report ~\cite{93}, the percentage of encrypted web traffic on the Internet has increased from around 50\% in 2014 to above 90\% in 2023. Therefore, these traditional methods are no longer applicable. In addition, DPI methods have raised concerns and discussions about user data privacy. On the other hand, the port-based traffic identification approach assumes that applications use well-known Transmission Control Protocol/User Datagram Protocol (TCP/UDP) port numbers assigned by the Internet Assigned Numbers Authority (IANA). Therefore, once an application does not follow the IANA standards, such as applying dynamic ports or encryption protocols such as peer-to-peer (P2P) protocols, the port-based traffic identification approach can no longer identify the traffic and the application ~\cite{niu2019heuristic}.

The second category represented by artificial intelligence technologies provides an effective solution. Traffic detection and classification models using machine learning or deep learning can effectively address user privacy and encryption issues. For example, network traffic can be effectively analyzed by side-channel features and malicious traffic can be identified. However, due to the unique characteristics of 5G networks, traditional methods of detecting malicious traffic used in previous generation network environments may not be suitable for 5G networks. For example, the distributed architecture, the huge amount of encrypted traffic data, and the large number of devices in 5G networks make it difficult to extract and process features and then detect malicious traffic in a traditional offline manner. In addition, 5G networks require real-time detection capabilities, which further increases the complexity and difficulty of the detection process.

As the use of 5G networks continues to grow, it is important to consider how adversaries may exploit them to launch attacks. For example, an enemy could use 5G networks to launch attacks on critical infrastructure such as the power grid, transportation systems, and medical facilities. To address these risks, it is important to understand the state-of-the-art (SOTA) technologies and their capabilities in malicious traffic detection over 5G networks. Therefore, the contribution of this paper is:

1. We delve into the security challenges and corresponding solutions that exist from 1G to 5G networks and then provide a comprehensive overview of 5G-related technologies, their respective security challenges, and potential solutions. 

2. We comprehensively review the current SOTA techniques for malicious traffic detection under AI and their applicability to 5G networks. The SOTA in 5G traffic detection is also analyzed and discussed. We will discuss the challenges and limitations of traditional detection methods if applied in 5G networks and explore emerging technologies. 

3. We proposed a set of evaluation criteria for 5G network traffic datasets and provided a summary and comparison of publicly available datasets.

4. We also propose and compare different incremental learning-based machine learning detection models for 5G network data that address the pain points of low robustness in current traffic detection.

The organization of the rest of this article is as follows: Section II presents the related surveys about 5G security. In Section III, we provide an introduction to the different detection techniques for malicious traffic detection and analysis. Then we discuss the challenges of AI based malicious traffic detection models in the 5G environment in Section IV. The analysis and comparison among the existing available 5G datasets are discussed in Section V. In Section VI, we present the setup of our experiments and conduct the performance evaluations. We conclude the paper in Section VII, by discussing the remaining challenges and future directions.

\section{\bf 5G Security}
In this section, we introduce the network security issue and solution from 1G to 5G and review related survey works and SOTA which discuss 5G network security. 

The 1G cellular system used analog signal processing, making it difficult to provide effective security services for 1G. There was no encryption mechanism technology in 1G. Eavesdropping, unauthorized access, and cloning were difficult to be effectively addressed. The 2G cellular network established four security services to be provided by the Global System for Mobile Communications (GSM): anonymity through the use of temporary identifiers, authentication that is confirmed by network operators, signal protection, and user data protection through encryption mechanisms. However, there are still challenges in resisting man-in-the-middle attacks and reverse engineering of encryption algorithms within 2G. The emergence of 3G was primarily to provide higher data rates than 2G, but it also upgraded the security architecture of the network by introducing the three security principles defined by 3GPP for 3G: Fully inherit the security features of 2G, overcome the limitations of 2G security, and add security features not present in 2G (Universal Mobile Telecommunications System (UMTS)). 

The 10th edition of 3GPP, i.e. LTE-Advanced, the 4G system, defined a set of security features: Access security, Network domain security, User domain security, Application domain security, and Security visibility and configurability. New security mechanisms were also designed for Machine Type Communications (MTC), Home eNodeBs or femtocells, and relay nodes. However, new security challenges also have emerged, such as eNBs being vulnerable to physical attacks, DoS attacks, and passive attacks on long-term keys.

Ahmad et al. ~\cite{ahmad2019security} studied the security status of 5G networks. Firstly, the authors conducted an in-depth analysis of the security history of previous generations of wireless cellular networks from 1G to 4G, addressing their corresponding security challenges and respective solutions. As 5G can connect almost all aspects of society, this development will bring a series of new threats and security vulnerabilities that will pose significant challenges to both present and future networks. For example, if critical power infrastructure is connected to 5G, an attack on the vulnerabilities of such critical infrastructure could cause catastrophic effects on the infrastructure and society that previous generations of networks cannot compare to. The authors then categorized the security challenges and solutions of 5G networks according to the access network, the backhaul network, and the core network. They also proposed security solution suggestions based on key technologies integrated into 5G, such as SDN, NFV, cloud computing, and multiple-input multiple-output (MIMO). 

Dutta and Hammad ~\cite{dutta20205g} also conducted a comprehensive survey on 5G and Beyond Security Challenges and took a systematic approach to 5G threat analysis and taxonomy. Unlike the previous paper, the limitation of the former was that it was published before 5G was widely applied, while this paper's analysis of 5G threats and challenges is more recent, closer to the present day. The authors summarized the unique security challenges of 5G and provided detailed descriptions, analysis, and taxonomy of these challenges.

Salahdine et al.~\cite{salahdine2023security}, on the other hand, classified the new attacks and threats faced by the key technologies used in 5G networks into three main paradigms: 1. Security Keys; 2. Use Case; and 3. Targeted Network Entity. The authors discussed security issues and challenges based on five perspectives: resources, mobility, routing, physical layer, and autonomous services. The security of 5G networks is emphasized as a critical issue, as it affects both user privacy and system security. Wireless communication networks have always had security vulnerabilities, making them vulnerable to various types of attacks. With the development of each generation of wireless networks, attacks have become more powerful as well. 

In the case of 5G, existing attacks have become even more potent, and new attacks will  emerge from the key 5G-driven technologies mentioned earlier. The authors then discussed 14 types of 5G security challenges in detail, such as Massive MIMO. Massive MIMO is one of the most promising technologies in 5G and beyond. However, due to the large number of antennas used to serve multiple users simultaneously, it is vulnerable to active and passive security attacks. Machine learning has been used to detect active attacks, but the massive amount of data generated by Massive MIMO systems results in significant data overhead and requires extensive training of the algorithms. Network slicing technology allows customized services to be provided in 5G networks. However, network slicing and user equipment as bridges can enable attackers to launch attacks from one slice to another. The authors also discussed 23 types of security attacks that threaten 5G key supporting technologies, such as DoS attacks, eavesdropping attacks, malware attacks, and hijacking attacks. The authors then proposed solutions to the 5G security risks, focusing mainly on the solutions that cryptography and artificial intelligence techniques can provide. For example, malicious monitoring systems based on machine learning algorithms to analyze network traffic or files. The combination of quantum concepts and cryptography also provides solutions for more secure communication.

The authors for \cite{dutta20205g}~\cite{salahdine2023security}~\cite{ahmad2019security} also addressed new privacy-related issues brought by 5G networks. Compared to previous generations of networks, the emergence of new architectures, technologies, and services in 5G networks will ultimately bring higher privacy risks to users and other stakeholders. For example, cloud computing allows users' personal data to be stored and processed in virtual environments, which is then shared with different stakeholders and operators. This makes it easier to lead to privacy breaches. 

Privacy challenges and attacks can be categorized into three types based on users: personal data, user identity, and location ~\cite{salahdine2023security}. Data attacks can be divided into attacks on data during transmission and illegal access to data in storage systems ~\cite{ahmad2019security}.  Potential attacks during transmission can be information modification, eavesdropping, and man-in-the-middle attacks ~\cite{tiburski2016security}. In this case, strong encryption techniques are needed to protect the data. For storage systems, a privacy by design approach needs to be adopted, which protects privacy in the design and manufacturing stages ~\cite{porambage2016quest}. Strong authentication and authorization mechanisms are also required to protect data from unauthorized access ~\cite{ikram2009simple}. There are also privacy issues regarding user location and identity ~\cite{kumar2018user}, which require various technologies such as cryptography and obfuscation ~\cite{masood2018incognito}. However, while strong encryption techniques protect data, malicious attacks using encryption technology also become more difficult to detect with advance warning. 

% The vast majority of current machine learning and deep learning models are complex black-box algorithms, which in turn further expands the overall attack surface. Machine learning and deep learning models can also be the target of attacks. For example, an attacker may have different levels of access to the model and the training data, and they can reduce the accuracy of the model by modifying the authenticity or integrity of the training data, or even render the model completely unable to detect attacks. In addition, attackers can exploit vulnerabilities in trained models to evade attacks that would be detected by the model. Furthermore, attackers can also use the powerful inferential and classification capabilities of machine learning to break 5G security solutions, such as decrypting user privacy.

\section{\bf Artificial Intelligence in legitimate and malicious traffic analysis and detection}

Artificial intelligence shows great potential in traffic analysis and detection within 5G network security. Analyzing massive amounts of data in 5G or monitoring security traffic in network transmissions requires a proactive, self-aware and adaptive intelligent system paradigm. The algorithms and techniques of artificial intelligence can assist analysts in analyzing traffic data without prior data or experience, and can continuously learn and update their own models. When the traffic data is too complex to be understood by traditional methods, machine learning can provide appropriate analytical advice within the allowable time. Therefore, network security may become one of the best application areas for artificial intelligence.

\subsection{\bf Traditional Solutions for Malicious traffic analysis and detection}

Prior to the widespread application of artificial intelligence technology and encrypted traffic, one of the most common methods for investigating network attacks was using Deep Packet Inspection (DPI). This technique performs a thorough inspection of the fields contained in the traffic packet. It enables the detection of anomalies in network traffic, as well as other important information useful for incident response, such as IP addresses, traffic types, attack durations, and other data helpful to security professionals in mitigating incidents. In theory, DPI has the ability to detect almost all intrusions by searching for packets and identifying content that matches known attacks.

Pimenta et. al ~\cite{pimenta2017cybersecurity} and El-Maghraby et. al ~\cite{el2017survey} provided detailed discussions on the DPI technology from both software and hardware perspectives, including its architecture and service requirements (such as memory efficiency, dynamic update, scalability, and others), as well as the detection techniques that stem from it, such as SNORT ~\cite{26}, Bro ~\cite{27}, and Linux L7-filter ~\cite{28}. The authors also emphasized the challenges and limitations faced by DPI. For example, since DPI systems need to be able to inspect all content of packets from any protocol, it is necessary to have a variety of rich and specific parsers. However, having rich and specific parsers also means a greater likelihood of software vulnerabilities. For example, the vulnerability CVE-2014-4174 was identified, which allowed attackers to execute arbitrary code in Wireshark ~\cite{sommer2016spicy}. The article ~\cite{pimenta2017cybersecurity} also points out that DPI can be applied for malicious purposes, such as interfering with network neutrality~\cite{bendrath2011end}, allowing government surveillance ~\cite{bendrath2009global}, and violating privacy ~\cite{fuchs2012implications}.

With the increase in bandwidth of 5G networks and the influx of large amounts of data into the internet, new network applications are emerging and existing ones are constantly being updated. This leads to the rapid expansion of traffic identification feature rules that need to be stored in the rule database, resulting in increased processing and storage costs for the system. Moreover, complete network payload analysis not only incurs high computational costs but may also involve user privacy disputes and data security breaches ~\cite{sun2016efficient}. As a result, DPI has encountered resistance during its development.

\subsection{\bf AI-based Malicious traffic analysis and detection}

% Machine learning has been applied to almost all engineering domains in the past decade. However, due to the lack of sufficient available and comprehensive network traffic datasets in the past, the combination of machine learning and network traffic analysis and detection has lagged behind other fields. Compared to other fields where there are lots of valuable training datasets available, network traffic datasets are relatively scarce, making it challenging to generate representative machine learning training sets. This is because the features of network traffic data are affected by a large number of environmental parameters and interference conditions. Moreover, AI-based network traffic detection systems require constantly learning and updating the latest data to perform accurate detection, which limits their efficiency and carries high risks.

In recent years, as the demand for privacy and data security has grown, enterprises have chosen to use encryption mechanisms to hide the context of their payloads. Given this trend, the amount of encrypted traffic data in global communication networks has sharply increased. However, the emergence of encrypted traffic is also being exploited by attackers to disguise their malicious behavior. 

DPI techniques, as mentioned above, are often powerless against  such encrypted malicious traffic ~\cite{niu2019heuristic}. The process of decoding and recognizing the payload of encrypted traffic by DPI is often slow and cannot support the detection of a large amount of traffic in an effective time frame. In addition DPI by the thorough inspection of the fields contained in the packets flowing in the network itself is contrary to the original purpose of the encryption mechanism applied to the network. Checking anomalies in network flows by decrypting the encrypted traffic is itself a kind of breach of user data privacy ~\cite{wang2023feature}. At this point, the application of machine learning technology has successfully solved the problem of DPI's privacy invasion. Detection and recognition of encrypted traffic based on non-decryption methods mainly rely on machine learning technology due to the ability for machine learning to be applied on non-payload based features such payload length, inter-arrival time of traffic packets, and flow duration. Benefiting from the rapid development of computer hardware in recent years, artificial intelligence has been easily applied in various fields and its performance has proven to be worth relying on in many fields.

Currently, research on malicious encrypted traffic detection focuses on feature extraction and the selection of machine learning or deep learning algorithms. The feature extraction in machine learning based malicious traffic detection can be summarized in two aspects: statistical features and protocol features, which we define as protocol-agnostic traffic feature and protocol specific traffic feature.

Protocol-specific features are features extracted from certain specific encryption protocols. These features are created by the specific properties of a particular encryption protocol so that such features can be considered as highly discriminatory features of encrypted traffic data. However, such features cannot be generalized across different protocols. For example, for certificate validity of TLS/SSL protocols and the average value of the public certificate keys cannot be extracted from the SSH protocol. In addition, extracting protocol-specific features usually requires the use of some specific extraction application tools and is time-consuming. 

Protocol-agnostic statistical traffic feature is based on the analysis and extraction of statistical information about network traffic. It consists of two granularities: packet layer, where the features are extracted from each packet, such as time to live per packet and payload length of each packet, and session layer, where the features are extracted over a complete network traffic flow session, such as flow duration, total payload length per session. The protocol-agnostic features extracted at both granularities can be analyzed separately or together. In addition, many authors classify protocol-agnostic statistical traffic features into categories that are suitable for their proposed models. For example, Bekerman et al.~\cite{bekerman2015unknown} divided traffic data into four groups (conversation window, flow, session, transaction) to extract protocol-agnostic features for model training. Bader et al.~\cite{inproceedings72} proposed a novel feature creation approach that extracts features from the first 32 traffic packets based on bidirectional packets, source, destination, handshake packets, and data transfer packets. Based on the attributes of features, Liu et al.~\cite{liu2019distance} grouped their extracted features into TCP/IP header features (e.g., payload header), time-based features (e.g., flow duration, average inter-arrival time of packets), length-related features (e.g., packet size and payload size), and packet variation features (e.g., TCP window change times). Wang et al.~\cite{wang2023feature} analyzed encrypted traffic sessions and packets to extract unique features that only encrypted traffic possesses. Generally, protocol-agnostic features do not have the limitations of specific protocol features. They are more robust and easier to extract. However, there are many types of them, and the extraction process often requires prior knowledge and consumes a lot of time. Both protocol-agnostic statistical traffic features and protocol-specific traffic features can be subjected to feature engineering to create the minimum, maximum, mean, median, standard deviation, and variance values of each numerical traffic feature and further expand the size of feature sets for model training.

After extracting and selecting a suitable feature set, the design of the algorithm is equally important. Zhang et al.~\cite{article67} proposed a cascade forest (CaForest) which is simplified with only cascade structure from the multi-grained cascade forest (gcForest) framework. The CaForest addresses the problem of detecting SSL/TLS encrypted malicious traffic from small and unbalanced training data. The authors characterize the traffic according to the SSL/TLS protocol, the traffic is partitioned into sessions based on 5-tuple information. Each session is then transformed into a two-dimensional traffic image that is used as input to the deep learning model. Their experiments demonstrate detection rates ranging from 6.87\% to 29.5\% higher on small and unbalanced datasets compared to other SOTA deep learning based methods. 

% This addresses the dataset criterion we proposed above about insufficient number of encrypted traffic and class imbalance of the traffic dataset. 

Bazuhair et al ~\cite{inproceedings70} also proposed an encoding method to convert selected features of Transport Layer Security (TLS)/Secure Sockets Layer (SSL) traffic into a two-dimensional image with data argumentation by Perlin noise. By processing the CTU-13 dataset, the final generated image-like data inputs were used to train a convolutional neural network (CNN) binary classification model and achieve a false negative rate of 0.40\% and a false positive rate of 5.60\%.

Liu et al.~\cite{liu2023nt} proposed a novel network traffic graph neural network model (NT-GNN) for Android malware detection based on network traffic graph. The authors proposed that some current malicious traffic detection ignores the complex structural relationships of network traffic and focuses only on the network traffic between endpoints. The authors proposed a network traffic graph neural network model (NT-GNN) which considers the node and edge aspects of the graph, capturing the connections between various traffic flows and individual traffic attributes. The model achieves 97\% accuracy in the CICAndMal2017 and AAGM datasets.

Zheng et al.~\cite{article69} pointed out that traffic classification methods contain statistical feature-based methods and graph-based methods. However, the statistical feature-based method focuses on the internal information of the network flows and the graph-based method focuses on the external connections between network flows. The authors considered both the statistical features of the network flows (internal information) and the structural information between them (external connections) to propose a graph convolutional network based encrypted malicious traffic detection (GCN-ETA). GCN-ETA consists of a modified GCN based feature extractor and a decision tree classifier to effectively improve the effectiveness and speed of encrypted malicious traffic detection. The experiment shows the method can perform 98\% higher accuracy, AUC and F1 scores, and achieve more than 1300 traffic flows per second.

Zhang et al.~\cite{inproceedings71} proposed an Efficientnet based encrypted malicious traffic detection by a transfer learning approach. The authors first pre-trained Efficientnet on Imagenet dataset, then omitted the manual feature extraction by experts and transferred it for training with a small number of encrypted traffic datasets. The model achieved 100\% detection accuracy and recall with only a small number of datasets in the experiment. However, the robustness of the model remains to be determined as the authors used a small traffic dataset.

Bader et al.~\cite{inproceedings72} proposed MalDIST based on the extension of the DISTILLER model ~\cite{ACETO2021102985} to encrypted malicious traffic detection and classification. The proposed framework for encrypted malware traffic detection and classification consists of several deep learning models, including 1D CNN, 2D CNN, and bidirectional GRU models. The proposed MalDIST achieved 99.7\% accuracy, precision, recall and F1 in experiments, outperforming seven other different algorithms.

Based on the TLS handshake and payload features, Ferriyan et al.~\cite{f8e214af22de4c1fa0ee9098706b1d95} proposed an encrypted malicious traffic detection (TLS2Vec). The proposed model does not need to wait for the traffic session to finish, which ensures privacy to a certain extent. LSTM and BiLSTM models are trained with words generated from selected features. TLS2Vec performs better from the CTU malware facility project dataset than Non-TLS2Vec, which uses neither symbols nor Word2Vec embedding. 

Meghdouri et al.~\cite{meghdouri2020cross} proposed a random forest classification model with a novel cross-layer feature representation method for TLS and Internet Protocol Security (IPSec) protocols. 100\%, 92.60\%, and 92\% F1 scores are achieved from three different public datasets. A support vector machine (SVM) and CNN based method for classification of malicious traffic for TLS encryption is proposed by Lucia and Cotton ~\cite{de2019detection}. The proposed 1D-CNN achieves 99.91\% accuracy and F1 score. The proposed SVM with radial basis function kernel achieves 99.97\% accuracy and F1 score.

An approach, TLARNN, is proposed by Liu et al.~\cite{liu2023spatial} which combines spatial-temporal features with a dual-attention mechanism. The TLARNN model contains 4 parts: spatial feature learning by using 1D-CNN and BiGRU, attention mechanism based on encrypted packet, flow temporal feature learning between encrypted streams, and attention mechanism based on flow. The approach provides rich encrypted traffic characteristics and insights for encrypted malicious traffic detection and classification.

\section{\bf Challenges of AI-based malicious traffic detection models in the 5G environment.}

The popularity and application of 5G networks have begun to show results~\cite{80}, and its higher speed, lower latency, higher capacity, better reliability, and lower energy consumption have received a lot of attention and expectations. In the field of AI-based malicious traffic detection, some research has also started to focus on malicious detection in the context of 5G networks.

Rezvy et al.~\cite{inproceedings44} proposed a deep auto-encoded dense neural network algorithm for intrusion detection in 5G and IoT networks. The proposed model combines an unsupervised pre-training layer with a supervised dense neural network for malicious attack classifications. The AWID dataset was used to evaluate the algorithm, and it achieved an overall accuracy of 99\% for detecting flooding, impersonation, and injection attacks.

Xie et al.~\cite{xie2021traffic} discussed how to classify traffic types, block malicious traffic, and effectively utilize sensor data in the context of 5G network slicing. Authors designed how to connect ZigBee devices to the home IoT to obtain sensor data such as temperature, humidity, home power, and smoke concentration in the home. Proposed Chi-square filter method achieves the highest accuracy and better overall performance in Moore Dataset ~\cite{liangjun2018correntropy}. However, the authors did not perform evaluation on their own collected 5G data but use Cambridge University's Moore dataset as a training test set for traffic classification.

Lam and Abbas~\cite{lam2020machine} proposed SDS (Software Defined Security) aiming to provide a proactive end-to-end defense for 5G networks using CNN model with NAS (Neural Architecture Search). Application attempts to provide an adaptive security solution for 5G networks. The authors selected the most recent IDS dataset available at the time, namely CICIDS2018, and then generated an optimized model for the specific dataset by using a NAS-based CNN model. The model achieved 100\% of benign traffic Recall and 96.4\% of anomalous traffic Recall.

Through a broad search of network traffic detection in 5G network architectures, many authors propose "best practices" for network monitoring architectures that remain to be determined. For example, in~\cite{lam2020machine}~\cite{inproceedings44}~\cite{xie2021traffic}~\cite{maimo2018self}~\cite{lei2022anomaly}, the authors proposed their novel approaches for 5G based traffic detection and analysis, respectively. However, none of the authors of the papers collected or used real traffic under real 5G networks for analysis. They invariably chose the simulated and past datasets for their performance evaluation.

The 5G network is different from previous generations of networks, as it is considered an extremely flexible infrastructure based on an architecture organized by different functional modules. Therefore, research on 5G cannot be simply considered as an accelerated upgrade of the 4G network. The challenges that have not been the focus of much previous research are being rapidly amplified by the characteristics of 5G and are becoming huge challenges that need to be addressed. Among these magnified challenges, we believe that the three most urgent challenges that need to be addressed are: high-performance processing of detection models, traffic data security and privacy, and high robustness of detection capabilities.

\paragraph{\bf \textit{High-performance processing of the detection models}}

The 5G network can provide higher network speeds and capacity, which means that the scale of network traffic will be larger. Therefore, AI models need to have high performance to process large amounts of incoming traffic data. Traditional offline training detection models should be able to achieve the same processing efficiency online. However, in most of the current known research, the detection speed and efficiency of models are not considered in experimental settings. 

Maimó et. al ~\cite{maimo2018self} proposed a self-adaptive architecture for deep learning based 5G network defense. The authors utilized deep learning to inspect network traffic by extracting features from the traffic. Additionally, the proposed framework could automatically adjust the configuration of the network defense architecture to effectively manage fluctuations in traffic. This allows for optimization of resource allocation at any given moment, as well as fine-tuning the performance and behavior of the analysis and detection process. This demonstrates the potential of their proposed model in 5G network detection, particularly in self-adapting the anomaly detection system based on real-time detection of the volume of network flows gathered from 5G, as well as optimizing resource consumption. 

Fu et. al ~\cite{fu2023detecting} proposed an ML-based real-time malicious traffic detection system, which is named HyperVision. The proposed model detects unknown patterns of encrypted malicious traffic by using an in-memory graph built on the aggregate of long and short traffic information interactions. The authors develop an information-theoretic analysis framework to show that the in-memory graph captures near-optimal traffic interaction information. In experiments conducted with 92 real-world attack traffic datasets, a detection throughput of at least 80.6 Gb/s is achieved with an average detection latency of 0.83 seconds. 

Zheng et al.~\cite{article69} proposed a GCN-based encrypted malicious traffic detection by using an improved feature extractor of GCN and a classifier using decision trees to improve the effectiveness and speed of encrypted malicious traffic detection. The method was experimentally implemented to detect more than 1300 traffic flows per second.

\paragraph{\bf \textit{Traffic data security and privacy}}

The demand for detecting encrypted traffic is greater in 5G networks than in previous generations of network due to the large volume of traffic and the possibility of more types of encrypted malicious traffic. This requires training encrypted malicious traffic detection models using larger amounts of network traffic data. Therefore, analyzing encrypted traffic without compromising data security and privacy protection is an important issue. Existing work, such as ~\cite{inproceedings71}~\cite{inproceedings72}~\cite{ACETO2021102985}~\cite{meghdouri2020cross}~\cite{fu2023detecting}, has proposed their respective viewpoints on this topic. 

HyperVision by Fu et al.~\cite{fu2023detecting} uses two different strategies to represent the interaction patterns generated by short and long flows, and aggregates the information of these flows to establish a memory graph. Using a proposed unsupervised graph learning method, the connectivity, sparsity, and statistical features of the graph are analyzed and trained for detection without any known attack-labeled dataset, thereby achieving identification of malicious encrypted traffic. Novel cross-layer feature representation for encrypted traffic is proposed by Meghdouri et al.~\cite{meghdouri2020cross} from 3 defined extracting flows: Application flows, conversation flows, and End-point flows. Encoding method ~\cite{inproceedings71} that converts protocol-specific encrypted traffic features into image-like data to conduct further encrypted malicious traffic analysis and detection. 784 payload bytes of each session and packet size, windows size, and interarrival time of the first 32 packets are proposed by the work of Aceto et al.~\cite{ACETO2021102985} for encrypted traffic analysis and detection. Bader et al ~\cite{inproceedings72} provide another feature extraction approach for encrypted traffic by grouping the first 32 traffic packets into 5 groups (bidirectional packets, source, destination, handshake packets, and data transfer packets) and computing more statistical features according to above groups. Works in ~\cite{wang2023feature} also proposed an encryption feature creation approach specifically for analyzing encrypted traffic. The approach considers both flow session and packets which can be viewed as the exclusive feature of encrypted traffic analysis. It overcomes the limitation that traditional protocol-agnostic traffic feature creation methods which are often unable to capture the unique characteristics of encrypted traffic. 

\paragraph{\bf \textit{High robustness of detection capabilities.}}

Current approaches of malicious traffic detection rely on the assumption that the model which learned from previous malicious traffic characteristics will be effective in identifying new and future malicious traffic. However, the types and forms of malicious traffic are constantly evolving, and new malicious traffic is constantly emerging. Traditional machine learning algorithms are actually difficult to adapt to these changes in real-time. For example, identifying encrypted malicious traffic generated by 0-day malware is hard to detect. Based on Wang et al.~\cite{wang2022machine}, significant differences in the distribution of data structures between different datasets have been demonstrated. For example, the differences in distribution between IoT and conventional device traffic data make detection models trained with conventional device traffic data perform poorly when faced with IoT traffic data. Therefore, detection models trained on a fixed training set, such as ~\cite{inproceedings70}~\cite{meghdouri2020cross}~\cite{yao2019identification}~\cite{zeng2019deep}, are susceptible to becoming outdated and at risk of losing their detection capabilities.

To address these issues, AI algorithms with certain generalization abilities, which can adapt to different types and forms of malicious traffic are needed. Incremental learning, as an online learning method, can gradually update and improve models with the continuous input of new data while maintaining model accuracy. This can to some extent address the issue of robustness in detecting malicious network traffic.

Incremental learning is the process of continuously updating and improving an existing model by adding new data and knowledge to the model. Unlike traditional machine learning methods, the goal of incremental learning is to enable machines to continue learning in a constantly changing environment and adapt quickly to new data and knowledge. By continually updating the model, it can adapt to new types and forms of malicious traffic, thereby improving the robustness of detection. 

In recent years, some researchers have begun to explore the application of incremental learning in the field of malicious traffic detection. They have attempted to use incremental learning algorithms to improve the robustness, accuracy, and efficiency of malicious traffic detection, while reducing computational and storage costs. For example, Xu et al.~\cite{xu2022self} proposed a self-evolving malware detection (SEMD) method that uses network traffic and incremental learning. It can learn new tasks without forgetting old tasks, and borrows the idea of knowledge distillation through the loss function. Experimental results show that the proposed method can identify both old and new tasks, overcoming the problem of network forgetfulness. An incremental learning based Support Virtual Machine (SVM) encrypted malicious traffic detection model is proposed by Lee et. al ~\cite{lee2020encrypted}. After using the stochastic gradient descent algorithm, the proposed model showed low false discovery rate and high off-line and on-line accuracy. In the following Section VI, we conducted and compared 4 different algorithms for encrypted malicious traffic detection based on incremental learning techniques, to try to address the robustness issues faced by most existing research works. In conclusion, if the three key issues mentioned above are not addressed and upgraded accordingly, 5G technology poses a risk of rendering the current detection model obsolete.

\section{\bf {Evaluation of publicly available network traffic datasets}}

Compared to other research areas in AI applications, the quantity of malicious traffic datasets is scarce, and many datasets cannot be made available to the public due to privacy issues, such as ~\cite{masood2018incognito}, or only partial information can be released, such as ~\cite{36}. Additionally, after anonymization, the structure of network traffic datasets may have changed and is no longer entirely trustworthy. A suitable dataset is crucial for AI-based detection models, but due to the scarcity of appropriate datasets, many research experiments still use outdated datasets. For example, in ~\cite{yadav2022intrusion}, the UNSW-NB15 dataset was used to train and test the model. The UNSW-NB15 ~\cite{moustafa2015unsw} is a dataset released in 2015. Rezvy et al.~\cite{inproceedings44} proposed a deep learning model to improve the detection performance of the Aegean WiFi Intrusion Dataset (AWID)~\cite{42}, which was also made publicly available in 2015.

To address these issues, we propose an evaluation standard for determining whether a network traffic dataset is suitable for machine learning. The proposed standard is particularly applicable to datasets for malicious traffic monitoring after the widespread adoption of 5G. To the best of our knowledge, this is the first evaluation standard for 5G network malicious traffic detection and analysis datasets, which will help researchers in this research domain to evaluate the quality of datasets. We believe that a dataset suitable for machine learning-based malicious traffic detection and analysis for 5G networks should meet most of following criteria:

1. Sufficient amount of traffic data with clear and detailed description

2. Sufficient amount of malicious traffic data

3. High variety of malicious traffic types

4. Ground Truth of the traffic

5. Contain real 5G traffic (for 5G malicious traffic detection) 

6. Contains encrypted traffic (for encrypted malicious traffic detection)

7. Traffic data is highly mineable

% First of all, the dataset should have sufficient network traffic data to enable adequate training and prediction of machine learning or deep learning algorithms. Compared to traditional machine learning, deep learning requires a larger amount of data. This is because deep learning models typically contain a large number of parameters and complex structures, which need to be optimized through a large amount of data during training. In addition, deep learning models typically have strong generalization capabilities, meaning they can make effective predictions and classifications on unseen data. To improve the model's generalization ability, more data is needed to train the model and prevent overfitting. The traffic dataset should have clear documentation, including essential information about the data collection phase, such as the duration of the entire data collection, types of malicious attacks, IP addresses of infected and normal hosts, etc. The dataset should have a variety of different types of malicious traffic activities to ensure diversity in the traffic data, such as Trojan, DoS, and Ransomware. Clear labeling is necessary, and an appreciated dataset should have clear documentation of labels for the captured traffic files. For example, in 5G-NIDD [40], each traffic flow is labeled with two layers of labels: one layer indicates whether the traffic is malicious or benign, and the other layer provides detailed labels for the type of malicious traffic (e.g., SYNFlood, SlowrateDoS). 

First of all, the dataset should have sufficient network traffic data to enable adequate training and prediction of machine learning or deep learning algorithms. The traffic dataset should have clear documentation, including essential information about the data collection phase, such as the duration of the entire data collection, types of malicious attacks, IP addresses of infected and normal hosts, etc. The dataset should have a variety of different types of malicious traffic activities to ensure diversity in the traffic data, such as Trojan, DoS, and Ransomware. Clear labeling is necessary, and an appreciated dataset should have clear documentation of labels for the captured traffic files. For example, in 5G-NIDD ~\cite{samarakoon20225g}, each traffic flow is labeled with two layers of labels: one layer indicates whether the traffic is malicious or benign, and the other layer provides detailed labels for the type of malicious traffic (e.g., SYNFlood, SlowrateDoS).

Furthermore, one of the best ways to share the traffic data is by providing the raw traffic which is commonly captured and stored as PCAP files. It has the highest potential for feature mining. It allows for the extraction of traffic features from various perspectives, such as using specific tools (e.g., ZeekIDS ~\cite{64}) to extract protocol-specific traffic features ~\cite{wang2022machine}, or mining desired statistical data or protocol-agnostic traffic features based on side channels. Many public datasets come in the form of pre-processed csv format that usually has cleaned, filtered, and extracted features such as ~\cite{36}. Such datasets greatly save time for users in processing the traffic data, and allow training and testing of AI models without requiring significant domain knowledge. However, if the required features for training the model are not included in such datasets, such as ~\cite{inproceedings53}, then the dataset cannot be used. 

Additionally, in training models for encrypted traffic, if the csv format feature dataset does not have labels indicating whether the features were extracted from encrypted or unencrypted traffic, and the source of the dataset does not provide the original traffic files (i.e., PCAP files), it becomes challenging to train models for encrypted traffic detection. Therefore, such datasets will lack many data mining capabilities compared to datasets that are in PCAP formats. Moreover, with the application of encryption mechanisms, many past network traffic datasets are no longer appliable for encrypted malicious traffic analysis and detection, such as ~\cite{moustafa2015unsw}~\cite{60}. 5G traffic specifically collected from 5G network environment is even rarer. Finally, only by ensuring the ground truth of the dataset can we guarantee the true validity of the model.

In order to find suitable and effective public datasets, we collected currently available public datasets and summarized them based on the aforementioned criteria to lay a good foundation and provide a basis for future research. We have compiled the dataset list in Table I.

\begin{table*}[h]
\centering
\caption{The summary of public traffic datasets}
\begin{tabular}{|l|l|l|l|l|l|l|l|l|l|}
\hline
No. & Dataset                                   & \begin{tabular}[c]{@{}l@{}}Year of \\ release\end{tabular} & C1 & C2 & C3 & C4 & C5 & C6 & C7 \\ \hline
1   & 5G-NIDD ~\cite{samarakoon20225g}                          & 2022                                                       & Y  & Y  & Y  & Y  & Y  & L  & L  \\ \hline
2   & 5GAD-2022 ~\cite{coldwell2022machine}             & 2022                                                    & Y  & Y  & Y  & Y  & Y  & L  & L  \\ \hline
3   & CIRA-CIC-DoHBRW-2020 ~\cite{MontazeriShatoori2020DetectionOD}             & 2020                                                       & Y  & N  & N  & Y  & N  & Y  & Y  \\ \hline
4   & MQTT-IoTIDS2020 ~\cite{inbook56}                  & 2020                                                       & Y  & Y  & Y  & Y  & N  & L  & L  \\ \hline
5   & IoT Encrypted Traffic ~\cite{inproceedings63}             & 2020                                                       & Y  & N  & N  & Y  & N  & L  & Y  \\ \hline
6   & IoT-23 Dataset ~\cite{54}                   & 2020                                                       & Y  & Y  & Y  & Y  & N  & L  & L  \\ \hline
7   & MIRAGE-2019 dataset ~\cite{8888137}              & 2019                                                       & Y  & N  & N  & Y  & N  & L  & L  \\ \hline
8   & UNSW NS 2019 ~\cite{inproceedings62}                      & 2019                                                       & Y  & Y  & Y  & Y  & N  & Y  & Y  \\ \hline
9   & bot-IoT ~\cite{article55}                          & 2019                                                       & Y  & Y  & Y  & Y  & N  & Y  & L  \\ \hline
10   & CICIDS 2017 ~\cite{inproceedings53}                      & 2017                                                       & Y  & Y  & Y  & Y  & N  & Y  & Y  \\ \hline
11  & CIC-AndMal 2017 ~\cite{inproceedings51}                  & 2017                                                       & Y  & Y  & Y  & Y  & N  & Y  & Y  \\ \hline
12  & VPNnon-VPN traffic dataset ~\cite{inproceedings52}       & 2016                                                       & Y  & N  & N  & Y  & N  & Y  & Y  \\ \hline
13  & FIRST 2015 ~\cite{50}                       & 2015                                                       & Y  & Y  & Y  & Y  & N  & L  & L  \\ \hline
14  & UNSW NS 2015 ~\cite{moustafa2015unsw}                     & 2015                                                       & Y  & Y  & Y  & Y  & N  & Y  & L  \\ \hline
15  & Aegean WiFi Intrusion Dataset ~\cite{42}    & 2014                                                       & Y  & Y  & Y  & Y  & N  & L  & L  \\ \hline
16  & Malware Capture Facility Project ~\cite{49} & 2013                                                       & Y  & Y  & Y  & Y  & N  & Y  & Y  \\ \hline
17  & CICIDS 2012 ~\cite{article48}                      & 2012                                                       & Y  & Y  & Y  & Y  & N  & Y  & L  \\ \hline
18  & CTU-13 ~\cite{47}                           & 2011                                                       & Y  & Y  & Y  & Y  & N  & L  & L  \\ \hline
19  & CAIDA DDoS 2007 ~\cite{57}                   & 2007                                                       & Y  & Y  & Y  & Y  & N  & L  & L  \\ \hline
20  & Webldent 2 Traces ~\cite{inproceedings46}                & 2006                                                       & Y  & Y  & Y  & Y  & N  & L  & Y  \\ \hline
21  & Kyoto Dataset ~\cite{36}                    & 2006                                                       & Y  & Y  & Y  & Y  & N  & L  & L  \\ \hline
22  & LBNL ~\cite{58}                             & 2004                                                       & Y  & Y  & Y  & Y  & N  & L  & L  \\ \hline
23  & KDD Cup 99 ~\cite{59}                       & 1999                                                       & Y  & Y  & Y  & Y  & N  & L  & L  \\ \hline
24  & DARPA ~\cite{60}                            & 1998                                                       & Y  & Y  & Y  & Y  & N  & L  & L  \\ \hline
\multicolumn{10}{l}{C1 refers Sufficient amount of traffic data with clear and detailed description; C2 refers Sufficient}\\
\multicolumn{10}{l}{amount of malicious traffic data; C3 refers High variety of malicious activity; C4 refers Ground Truth}\\
\multicolumn{10}{l}{of the label of the traffic; C5 refers Contain 5G traffic (for 5G malicious traffic analysis and}\\
\multicolumn{10}{l}{ detection); C6 refers Whether encryption is applied to the dataset (for encrypted malicious traffic }\\
\multicolumn{10}{l}{ analysis and detection);C7 refers Traffic data is highly mineable. Y means Yes; N means No;}\\
\multicolumn{10}{l}{ L means Limit.}\\
\end{tabular}
\end{table*}

Through the above analysis, we can find that most public datasets have their own advantages and disadvantages, and generally only meet a few of the above criteria. In particular, there are few datasets that contain encrypted malicious traffic. There are even fewer datasets that can be used to represent real 5G networks.

We next analyze two approaches to network traffic data generation: real traffic collection and simulated traffic generation, both of which have their own advantages and disadvantages ~\cite{wang2022machine}. For real traffic collection, it guarantees the authenticity of the network traffic in the dataset, but there is often a lot of redundant data as well. Also the frequency of real network attacks is very low compared to legitimate network traffic, which leads to an imbalance between legitimate and malicious traffic. For simulated traffic generation, it can contain more malicious action types, which can ensure a high variety of malicious activities in the dataset. In addition, legitimate and malicious traffic of this type of dataset can be more balanced and more suitable for machine learning training. However, since this traffic is simulated, it may have imperceptible differences from the real network traffic data and cannot fully represent the network traffic conditions at this time. Unfortunately, there are no proven and well-recognized datasets for this domain-specific problem. Thus, researchers often use their own private datasets or select from some available public datasets. This poses a challenge for making fair performance comparisons in these existing works. 

After obtaining a suitable dataset, a series of pre-processing on the data is also required to make the data input suitable for training the constructed algorithm model. The first step is to remove irrelevant network packets from the raw traffic data that are not suitable for traffic analysis and detection studies, such as Address Resolution Protocol (ARP) or Internet Control Message Protocol (ICMP) packets. Then, duplicate, corrupted, and incorrect traffic that has not been fully captured also needs to be removed. Next, selected traffic features are extracted according to the needs of the algorithmic model and processed into the required data input structures, such as 2D image-like traffic feature data input format ~\cite{article67}~\cite{inproceedings68}.

In addition, for non-quantitative data, we need to quantize it, such as through one-hot encoding. For numerical features, we need to normalize them to minimize the redundancy of relational tables and avoid redundant database anomalous attributes. Class balance also requires to be taken into account. Although the proportion of malicious traffic in real network traffic is very small, we need to avoid the problem of imbalanced data in model training. The main reason is that an imbalanced dataset can cause the model to tend to classify test samples into the more numerous legitimate traffic category, while ignoring the less numerous malicious traffic category. This can ultimately result in our detection model being unable to accurately classify the type of traffic. Re-sampling and other methods can increase the frequency of the rare class appearing in the training data. Additionally, adjusting class weights can also solve this problem, such as assigning higher weights to malicious traffic to make the detection model pay more attention to the less numerous malicious traffic categories.

\section{\bf {Methodology and Experiment Design}}
\subsection{\bf {Incremental Learning Method}}
Most SOTA researches use traditional offline learning methods to train their proposed models. Since the training set and test set belong to the same data source, such offline trained models can achieve very high detection accuracy, such as above 98\% accuracy. However, with the popularity of 5G, a large number of new types of malicious traffic and data distribution structures that are not included in the offline model training set will appear frequently in the network. ~\cite{wang2022machine} has proved that once a detection model is built using traditional offline learning approaches and is applied to real-time detection, it has low performance for new emerging traffic data types. 
 
Incremental learning effectively solves four pressing issues faced by traditional learning approaches:

1. Data growth issue: After the popularity of 5G, the data growth problem deserves to be paid more attention. The traditional learning approach requires all data to be learned and trained at once. This may lead to insufficient computational and storage resources for large-scale and high-latitude data, especially in deep learning. The emergence of incremental learning enables models to be updated gradually, solving the problem of loading huge amounts of train set data at once.

2. Resource utilization issue: Without using incremental learning, if the offline trained model wants to generate classification for newly emerged traffic data distributions, the whole model needs to be retrained offline again. However, the training set used for retraining may only have a small fraction of the data changed. The training set will also need to be continuously updated for retraining. This can result in a waste of resources. Incremental learning can leverage previously learned knowledge and fine-tune it for new data to achieve more efficient utilization of resources.

3. Model robustness issue: The traditional learning approach learns once and the model lacks the ability to analyze new incoming data. Incremental learning, on the other hand, allows new incoming data to be learned continuously. This allows the model to continuously improve its robustness.

4. Privacy and security issue: Traditional learning approaches require all data to be loaded into the model at once for training, which raises the privacy concern of data leakage. However, incremental learning updates only the new incoming data, reducing the risk of data leakage.

Therefore, in experiment 1, we perform a comparison between traditional learning and incremental learning to evaluate whether the incremental learning method can better address the robustness issue. Online batch learning will be applied in this experiment by updating models iteratively using batches of new incoming data. The updates are based on the gradient of the objective function with respect to the new incoming data. Then we adjust batch size to manage the trade-off between computational efficiency and model performance. After that, we consider the catastrophic forgetting issue that incremental learning may encounter. Catastrophic Forgetting refers to the phenomenon that when a model learns new tasks or new data, it may forget the previously learned task or data. In Incremental Learning (IL), the model needs to continue learning on the new data. The new information may update the parameters of the model, thus overwriting or destroying the old information. Thus, in experiment 2, we analyze and try to find out an optimal trade-off between models' forgetting rate and detection accuracy on new emerging data.

\subsection{\bf Dataset Selection and Analysis}
With the introduction of a plethora of new connections, features and services, the multitude of services and technologies included in 5G technologies make modern communication networks inherently complex and sophisticated. As a result, the implementation of real-time, proactive and adaptive security mechanisms throughout the network will be an important part of 5G and future communication systems. Therefore, the large amount of data collected from real 5G networks will play an important role in the training of AI/ML models to identify and detect malicious traffic. Samarakoon et. al ~\cite{samarakoon20225g} proposed 5G-NIDD, which is a fully labeled dataset built on a functional 5G test network. It contains eight intrusion types along with benign network traffic. which provides a very useful aid for our experiments.

The accuracy and efficiency of ML-based intrusion detection depends heavily on the quality of the dataset, and how close the behavior of the data is to the behavior of real network scenarios. Recent datasets, such as CICIDS2017, UNSW-NS15, and CTU-13, demonstrate some features that have been successfully applied to ML-based intrusion detection research. However, the behavior and uses of 5G and above networks mentioned in the section above are far from the testbeds or simulation platforms used to create these existing datasets. In addition, there are relatively few publicly available intrusion detection datasets from real mobile network operators in 5G networks~\cite{samarakoon20225g}. Thus, collecting a dataset that consists entirely of real-world 5G environments is crucial for machine learning or deep learning detection models. 

5G-NIDD is a fully labelled dataset built on 5G Test Network Finland ~\cite{95} (5GTN) is presented by Samarakoon et. al ~\cite{samarakoon20225g}. 5GTN is an open and evolving innovation ecosystem supporting 5G and beyond technology research. It provides an advanced platform for developing cutting-edge solutions, services, systems, and products in the fields of 5G and beyond, AI, and cyber security. It employs a range of measurement and monitoring tools across different network areas, aiming to offer application developers a high-quality test network that meets carrier standards~\cite{piri20165gtn}. For dataset creation, the authors selected the University of Oulu site within the 5GTN. The data collection is processed on two different days and captured both attack and benign traffic passing through the network at the two base stations.

The 5G-NIDD dataset is provided in several different versions, including packet-based and flow-based formats, which greatly increases the mineability of the dataset. Data for each attack session from two base stations are provided separately in pcapng format. The same files are also available in packet-based pcapng, argus, and csv feature formats after the General Packet Radio Service Tunnelling Protocol (GTP) layer is removed. The flow-based files with the GTP layer removed are merged together to create a file containing all attacks from both base stations. Table II is the statistical summary of the dataset.

\begin{table}[h]
\centering
\caption{The statistical analysis of the dataset}
\begin{tabular}{|c|c|c|c|c|}
\hline
\multicolumn{1}{|c|}{No.} & \multicolumn{1}{c|}{\begin{tabular}[c]{@{}c@{}}Type of\\ Traffic\end{tabular}} & \multicolumn{1}{c|}{\begin{tabular}[c]{@{}c@{}}Base\\ Station\end{tabular}} & \multicolumn{1}{c|}{\begin{tabular}[c]{@{}c@{}}No. of \\ Flow Session\end{tabular}} & \multicolumn{1}{c|}{\begin{tabular}[c]{@{}c@{}}Malicious\\ or\\ Benign\end{tabular}} \\ \hline
\multirow{2}{*}{1} & Legitimate traffic        & BS1          & 406959              & Benign           \\ \cline{3-5} 
                   &                                 & BS2          & 70778               & Benign           \\ \hline
\multirow{2}{*}{2} & UDPFlood       & BS1          & 175811              & Malicious        \\ \cline{3-5} 
                   &                                 & BS2          & 281529              & Malicious        \\ \hline
\multirow{2}{*}{3} & HTTPFlood      & BS1          & 76121               & Malicious        \\ \cline{3-5} 
                   &                                 & BS2          & 64691               & Malicious        \\ \hline
\multirow{2}{*}{4} & SlowrateDoS    & BS1          & 36092               & Malicious        \\ \cline{3-5} 
                   &                                 & BS2          & 37032               & Malicious        \\ \hline
\multirow{2}{*}{5} & TCPConnectScan & BS1          & 10022               & Malicious        \\ \cline{3-5} 
                   &                                 & BS2          & 10030               & Malicious        \\ \hline
\multirow{2}{*}{6} & SYNScan        & BS1          & 10019               & Malicious        \\ \cline{3-5} 
                   &                                 & BS2          & 10024               & Malicious        \\ \hline
\multirow{2}{*}{7} & UDPScan        & BS1          & 7887                & Malicious        \\ \cline{3-5} 
                   &                                 & BS2          & 8019                & Malicious        \\ \hline
\multirow{2}{*}{8} & SYNFlood       & BS1          & 4792                & Malicious        \\ \cline{3-5} 
                   &                                 & BS2          & 4929                & Malicious        \\ \hline
\multirow{2}{*}{9} & ICMPFlood      & BS1          & 613                 & Malicious        \\ \cline{3-5} 
                   &                                 & BS2          & 542                 & Malicious        \\ \hline
\end{tabular}
\end{table}

\subsection{\bf Experiment Setup and Performance Evaluation}

The experiment running: Intel(R) Core(TM) i7-10700K CPU @ 3.8GHz 64.0GB of RAM. In order to construct our models, scikit-learn, Pytorch, and Avalanche libraries for Python are used. When evaluating the performance of detection models, accuracy, F1, precision, and recall are selected to provide a more comprehensive analysis of performance results. We have also taken into consideration the issue of catastrophic forgetting in incremental learning. In incremental learning, the model learns new tasks by continuously receiving new data and updating itself. However, this process can lead to forgetting of previously learned tasks. The main reason for this is that when updating the model with new data, the existing model parameters are modified, which can disrupt or overwrite the representations of the old tasks. Thus, the forgetting rate in incremental learning is also calculated and considered in the following experiments to analyze the catastrophic forgetting in each model.

1. Accuracy: The proportion of all classified samples correctly classified by the classification model. True positive(TP) is the number of malicious traffic classified as malicious. True negative(TN) is the number of benign traffic classified as benign. False positive(FP) is the number of benign traffic classified as malicious. False negative(FN) is the number of malicious traffic classified as benign.
\[Accuracy =\frac{TP+TN}{Total samples}\]

2. Recall: It is the proportion of the number of samples correctly predicted to be positive by the classifier to the total number of samples actually positive. 
\[Recall =\frac{TP}{TP+FN}\]

3. Precision: It is the proportion of the number of samples predicted as positive by the classifier that are actually correct.
\[Precision = \frac{TP}{TP+FP}\]

4. F1 Score: Both the precision and recall of the classifier are considered. The F1 score ranges from 0 to 1, with higher values indicating better performance of the classifier.
\[F1score =2 *\frac{precision * recall}{precision + recall}\]

5. Forgetting Rate: Forgetting rate is an evaluation matrix that is used for evaluating the degree of catastrophic forgetting. During incremental learning, a model may forget previously learned data or information as it learns new data or information. The higher the forgetting rate, the much of previously learned information has forgotten.
\[Forget\_rate =\frac{Acc (before\_IL) - Acc (after\_n\_IL)}{Acc (before\_IL)}\]

\subsection{\bf Experiment 1: Performance Analysis of the different incremental learning algorithms}
Three commonly used algorithms for incremental learning are selected in experiment 1. They are SVM, logistic regression, and perceptron. We would like to achieve two experiment objectives by conducting this experiment:

1. Performance comparison between traditional learning method and incremental learning method for 5G network traffic data.

2. Analyze the performance of different incremental learning algorithms.

We conduct a series of experiments (1A and 1B) to compare the models built by Traditional Learning and Incremental Learning. We choose 3 different incremental learning algorithms (SVM, logistic regression, and perceptron) to select the optimal model. The selected feature used in this experiment is shown in Table III. It is important to note these set of features are protocol-agnostic statistical features that have been introduced in Section III. We have chosen to use such features as they are applicable to all types of network traffic including that in the 5G network. Such statistical features are able to capture the unique data value distribution patterns of 5G network traffic as well. Further research could be done in the future to investigate if certain features are more important to 5G networks or if there are features that can be extracted which are unique to 5G networks.

After feature selection, the numerical ranges of different features may vary greatly. Thus, data normalization needs to be applied. Data normalization can normalize the traffic data to a range of zero to one to reduce data redundancy. The normalized dataset enhances the integrity and efficiency of model training.

\begin{table}[h]
\centering
\caption{Feature Set Selection}
\begin{tabular}{|c|l|}
\hline
No. & Feature Name                                        \\ \hline
1   & Flow Duration                                       \\ \hline
2   & Time to live value of forward traffic               \\ \hline
3   & Time to live value of backward traffic              \\ \hline
4   & Total number of packet in the flow                  \\ \hline
5   & Packet number of forward traffic                    \\ \hline
6   & Packet number of backward traffic                   \\ \hline
7   & Total Bytes of the flow                             \\ \hline
8   & Total Bytes of forward traffic                      \\ \hline
9   & Total bytes of backward traffic                     \\ \hline
10  & Time Differece between packets                      \\ \hline
11  & Mean value of packet size from forward traffic      \\ \hline
12  & Mean value of packet size from backward traffic     \\ \hline
13  & Total payload size of the flow                      \\ \hline
14  & Payload size of forward traffic                     \\ \hline
15  & Payload size of backward traffic                    \\ \hline
16  & Loss rate of the flow                               \\ \hline
17  & Loss rate of forward traffic                        \\ \hline
18  & Loss rate of backward traffic                       \\ \hline
19  & Windows size of forward traffic                     \\ \hline
20  & Windows size of backward traffic                    \\ \hline
21  & Round trip time for TCP                             \\ \hline
22  & Time difference between SYN and ACK                 \\ \hline
23  & Time difference between ACK and packets             \\ \hline
24  & Flow rate of the flow                               \\ \hline
25  & Flow rate from the source address                   \\ \hline
26  & Flow rate from the destination address              \\ \hline
27  & The data transfer rate from the source address      \\ \hline
28  & The data transfer rate from the destination address \\ \hline
\end{tabular}
\end{table}

In order to simulate a real network traffic monitoring environment, we designed the dataset and training process as follows. The 5G-NIDD dataset contains 728316 samples for base station 1 (BS1) data and 487574 samples for base station 2 (BS2) data. Two separate base stations have different traffic data patterns. We perform two sets of experiments that involve using different sets of data for training and testing the models. The purpose of this is to first train an offline AI model with BS2. Part of the traffic in BS1 is used as new incoming data to test the generalization and robustness of the offline AI model. The rest of the data in BS1 is used to apply the incremental learning method to update the model so as to show the improvement of the model after incremental learning. Based on the design, the dataset is split into 4 different sub-datasets: offline learning training and test set, and incremental learning training and test set. The first step is to train the offline model, we use the 90\% data from BS2. In the table IV, it is shown the 438816 samples from BS2 are used to train the offline model. Then, we use the remaining 10\% samples from BS2 as the offline test set to evaluate the offline model. Next, we require samples for batch incremental learning to continuously update the model. We split the data from the BS1. In this case, BS1 is split into 9:1. Thus, 655484 samples are used as incremental learning training set for online batch learning where the batch size we set is 10,000. The batch size can be set in a manner to manage the trade-off between computational efficiency and model performance. Finally, the remaining 72832 samples from BS1 (incremental learning test set) and 48758 samples from BS2 (offline learning test set) are used to evaluate the detection and robust performance of both the offline model and incremental learning model. 

\begin{table}[h]
\centering
\caption{Dataset design for experiment}
\begin{tabular}{|c|c|c|c|c|}
\hline
\begin{tabular}[c]{@{}c@{}}Data \\ Source\end{tabular}        & \begin{tabular}[c]{@{}c@{}}Total \\ No. of \\ Samples\end{tabular} & \begin{tabular}[c]{@{}c@{}}Offline \\ Model \\ Training \\ Samples\end{tabular} & \begin{tabular}[c]{@{}c@{}}Incremental \\ Learning \\ Training \\ Samples\end{tabular} & \begin{tabular}[c]{@{}c@{}}Model \\ Evaluation \\ Samples\end{tabular} \\ \hline
\begin{tabular}[c]{@{}c@{}}BS2 \\ (Offline data)\end{tabular} & 487574                                                             & 438816                                                                          & -                                                                                      & 48758                                                                  \\ \hline
\begin{tabular}[c]{@{}c@{}}BS1\\ (Emerging data)\end{tabular} & 728316                                                             & -                                                                               & 655484                                                                                 & 72832                                                                  \\ \hline
\end{tabular}
\end{table}

Experiment 1A Performance results including Accuracy, F1 score, Precision, and Recall are recorded in Table V, Figures 1, 2, and 3. After the training and evaluation are completed, two sets of evaluation results are contained in Table V. The result indicates that all three models achieve high accuracy above 97\% and above 98\% F1 score when offline models are evaluated with the offline test set (BS2 data). However, when the offline models test new incoming data (BS1 data), their accuracy drops to 57-60\%, and the F1 scores drop to 67-69\%. The AUROC score of the Perceptron model decreases from 98.32\% to 64.56\%. Similarly, the AUROC score of SVM drops to 61.74\% and the Logistic Regression model drops to 61.65\%. This highlights the issue of robustness in offline AI models.

In the next experiment 1B, incremental learning is applied to the above three offline trained models. The performance of models under new incoming data before and after incremental learning (last batch) is recorded in Table VI, Figures 1, 2, and 3. The experimental results show that the three models have a significant improvement in the detection of new emerging test data after receiving incremental learning. For instance, the Perceptron model after incremental learning can achieve 95.72\% accuracy, which is 35.36\% higher than the 60.36\% accuracy before it is applied to incremental learning. The F1 score of Perceptron model also increases from 68.68\% to 95.05\%. All three models achieve above 92\% AUROC scores after incremental learning. According to the experimental results, we find that the SVM model has the most significant improvement in accuracy, F1 score, and AUROC score after using incremental learning. The time cost of each model is also recorded in Table VII.

% Experiment 1 Performance results including Accuracy, F1 score, Precision, and Recall are recorded in Table IV. The experimental results based on different selected algorithms are also plotted in Fig 1, 2, and 3 to conduct performance comparison.}

% In experiment 1A, 90\% samples of Base Station 1 are used for offline model training in both traditional learning and incremental learning methods. The 487574 samples of Base Station 2 are split into two separate sets. The first set which contains 90\% Base Station 2 data are used for further training in the incremental learning method. The remain samples of Base Station 2 are then used for testing the models in both methods to evaluate robust performance with new emerging data. Experiment 1A Performance result including Accuracy, F1 score, Precision and Recall are recorded in Table IV. In experiment 1B, 487574 samples of Base Station 2 are purely used for offline model training in both traditional learning and incremental learning methods. The 728316 samples of Base Station 1 are split into two separate sets. The first set which contains 90\% Base Station 1 data are used for further training in the incremental learning method. The remain samples of Base Station 1 are then used for testing the models in both methods to evaluate robust performance with new emerging data. Experiment 1B Performance result including Accuracy, F1 score, Precision and Recall are recorded in Table V. The experimental results based on different selected algorithms are also plotted on Fig 1, 2, and 3 to conduct performance comparison.}

\begin{table*}[]
\centering
\caption{Experiment 1A: Experiment results of trained offline models before incremental learning}
\begin{tabular}{|c|c|c|c|c|c|c|c|}
\hline
No & Algorithms & Before Incremental learning & Accuracy & F1 & Precision & Recall & AUROC \\ \hline
\multirow{2}{*}{1} & SVM & Offline testset & 97.77 & 98.72 & 97.50 & 99.96 & 92.25 \\ \cline{3-8} 
& & Incoming testset & 57.08 & 67.15 & 50.56 & 99.97 & 61.74 \\ \hline
\multirow{2}{*}{2} & \multirow{2}{*}{Logistic Regression} & Offline testset & 97.86 & 98.77 & 97.57 & 99.99 & 92.49 \\ \cline{3-8} 
& & Incoming testset & 56.97 & 67.11 & 50.50 & 99.99 & 61.65 \\ \hline
\multirow{2}{*}{3} & \multirow{2}{*}{Perceptron} & Offline testset & 99.10 & 99.48 & 99.54 & 99.42 & 98.32 \\ \cline{3-8} 
& & Incoming testset & 60.36 & 68.68 & 52.58 & 98.99 & 64.56 \\ \hline
\multicolumn{8}{l}{Selected models trained with BS2 offline, BS1 data is viewed as new incoming traffic}\\
\end{tabular}
\end{table*}

\begin{table*}[]
\centering
\caption{Experiment 1B: Experiment results of incoming test data with models before and after incremental learning}
\begin{tabular}{|c|c|c|c|c|c|c|c|}
\hline
No  & Algorithms  &   Incoming Testset   & Accuracy & F1    & Precision & Recall & AUROC \\ \hline
\multirow{2}{*}{1} & SVM & before Incremental learning
& 57.08     & 67.15 & 50.56 & 99.97    & 61.74   \\ \cline{3-8} 
                   &                                         & after Incremental learning
 & 93.29     & 92.38 & 92.09 & 92.67    & 93.22  \\ \hline
\multirow{2}{*}{2} & Logistic Regression    & before Incremental learning
 & 56.97    & 67.11 & 50.50  &99.99   & 61.65  \\ \cline{3-8} 
                   &                                         &after Incremental learning
 & 92.48    & 91.43 & 91.49  &91.38   & 92.36  \\ \hline
\multirow{2}{*}{3} & Perceptron             & before Incremental learning
 & 60.36    & 68.68 & 52.58   &98.99  & 64.56   \\ \cline{3-8} 
                   &                                         & after Incremental learning
 & 95.72    & 95.05 & 96.36   &93.78  & 95.51  \\ \hline

\end{tabular}

\end{table*}

\begin{table}[]
\caption{Time Cost of each model}
\begin{tabular}{|l|l|l|}
\hline
Algorithms          & \begin{tabular}[c]{@{}l@{}}Time Cost of Traditional \\ Offline Learning\end{tabular} & \begin{tabular}[c]{@{}l@{}}Average Time Cost of \\ Incremental Learning \\ per Batch\end{tabular} \\ \hline
SVM                 & 368ms                                                                                & 4ms                                                                                            \\ \hline
Logistic Regression & 350ms                                                                                & 5ms                                                                                            \\ \hline
Perceptron          & 323ms                                                                                & 4ms                                                                                            \\ \hline
\end{tabular}
\end{table}

% \begin{table*}[]
% \centering
% \caption{\textcolor{red}{Experiment 1B} Performance Result}
% \begin{tabular}{|c|c|c|c|c|c|c|}
% \hline
% No                 & Algorithms                              & Learning Method      & Accuracy & F1    & Precision & Recall \\ \hline
% \multirow{1} & \multirow{SVM} & Traditional Learning & 79.28    & 88.31 & 85.54     & 91.27  \\ \cline{3-7} 
%                    &                                         & Incremental Learning & 86.41    & 92.65 & 86.32     & 99.99  \\ \hline
% \multirow{2} & \multirow{Logistic Regression}    & Traditional Learning & 62.27    & 75.84 & 84.13     & 69.04  \\ \cline{3-7} 
%                    &                                         & Incremental Learning & 86.60    & 92.75 & 86.53     & 99.93  \\ \hline
% \multirow{3} & \multirow{Perceptron}             & Traditional Learning & 23.47    & 32.14 & 67.14     & 21.12  \\ \cline{3-7} 
%                    &                                         & Incremental Learning & 86.38    & 92.64 & 86.32     & 99.95    \\ \hline
% \multicolumn{7}{l}{Selected models trained with BS1 offline, incremental learning and testing with BS2}\\
% \end{tabular}
% \end{table*}

\begin{figure}
\centerline{\includegraphics[width=21pc]{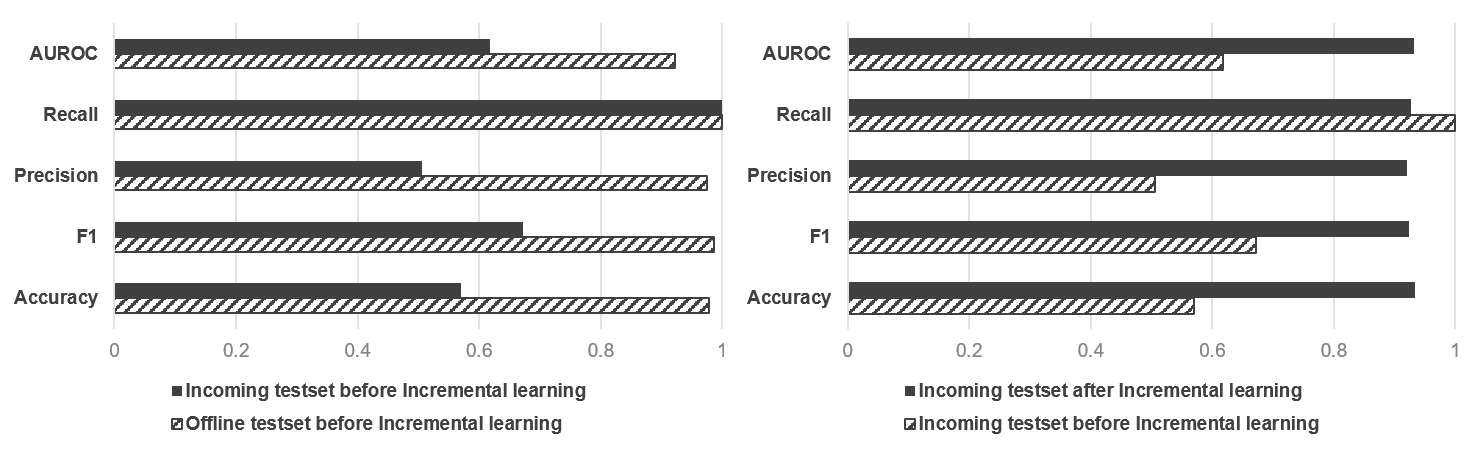}}
\caption{SVM performance comparison between traditional learning and incremental learning.}
\end{figure}

\begin{figure}
\centerline{\includegraphics[width=21pc]{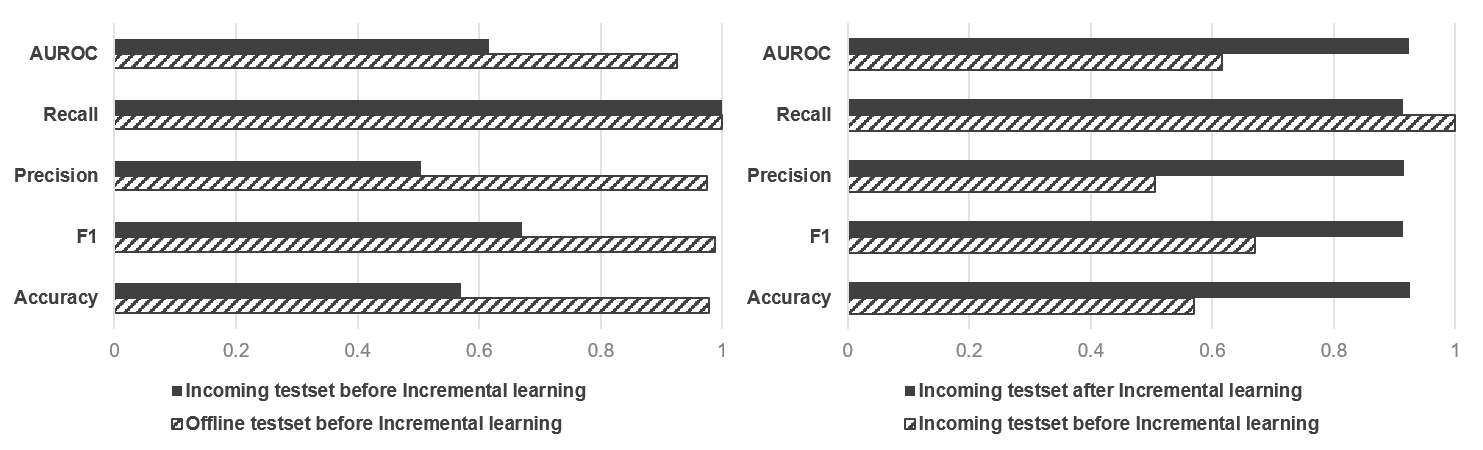}}
\caption{Logistic regression performance comparison between traditional learning and incremental learning.}
\end{figure}

\begin{figure}
\centerline{\includegraphics[width=21pc]{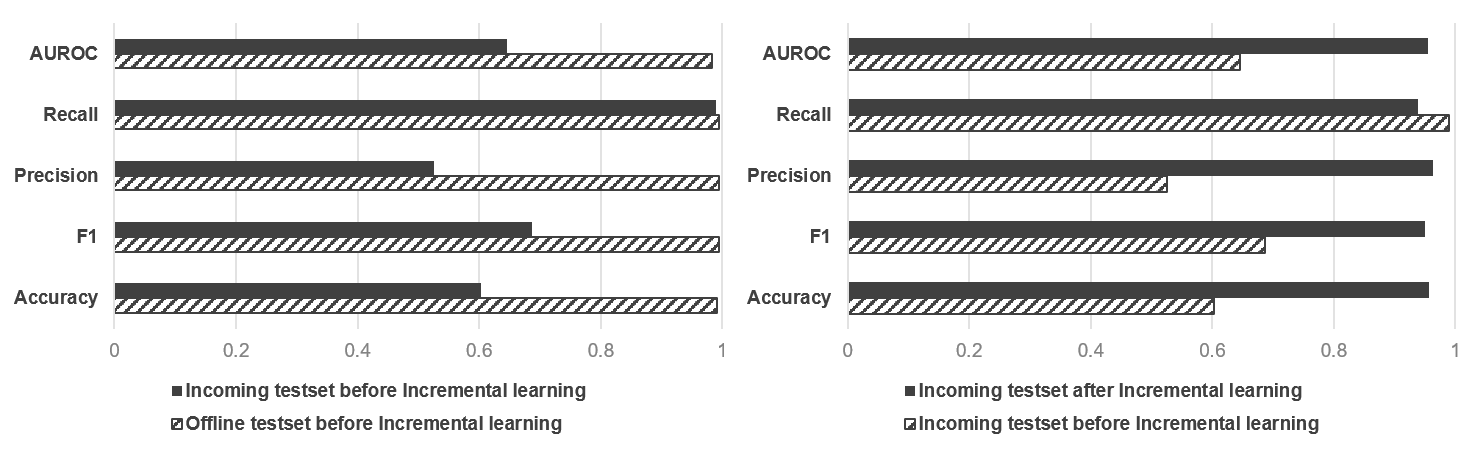}}
\caption{Perceptron performance comparison between traditional learning and incremental learning.}
\end{figure}

% \textcolor{blue}{Refer to the performance result in Table IV, V and Figure 1, 2, 3, the model incorporating incremental learning is stronger in terms of accuracy, F1 score,and AUROC score than the same algorithm using the traditional learning approach. For example in Table IV, we can observe that the accuracy of the SVM model based on Traditional Learning is 37.21\% which increased to 71.95\% after Incremental Learning was introduced. The Recall of the SVM model based on Traditional Learning is 54.45\% which increased to 97.81\%. Similarly in Table V results where a difference training set for Traditional Learning and Incremental Learning is used, the accuracy of logistic regression increased from 62.27\% to 86.60\%. }According to the experimental results, we find that the Perceptron model has the most significant improvement in each evaluation metrics after using incremental learning.

However, the above experiments are not yet sufficient to provide a conclusion on which model is most suitable for application to network traffic detection. One part of incremental learning that is still very much worth considering is the catastrophic forgetting issue. In experiment 2, we will further analyze the batch online learning we applied in experiment 1 and also apply different incremental learning approaches to address the forgetting rate.

\subsection{\bf Experiment 2: Performance Analysis for addressing catastrophic forgetting issue}

The objective of conducting experiment 2 is to analyze the relationship and trade-off between the model's forgetting rate and detection accuracy on new emerging data. In experiment 1, we applied an incremental online batch learning strategy to address the robustness issue of the offline model. However, the catastrophic forgetting issue is not considered in Experiment 1. In each batch round of the incremental batch learning process, each model can be evaluated with the same original offline test data to record the change of forgetting rates gradually. The formula of forgetting rate that has been introduced in Section VI.C. Simultaneously, new incoming test data is tested to show the change of detection performance in each batch round as well. In experiment 2A, the forgetting rate and detection performance of each model in each batch learning round are plotted in Fig 4 to 9. 

Figures 4,6, and 8 are plotted according to ascending order of batch number. They are used to analyze the relationship between forgetting rate and detection performance as the batch round number increases. Each figure has two graphs, The top one is the plot of model accuracy with respect to ascending order of batch number. Then, the below graph is the plot of the change of forgetting rate with respect to the ascending order of batch number.

Figure 5, 7, and 9 are plotted according to ascending order of accuracy performance. The top graph in each figure plots the accuracy of the model in each batch as a continuously rising line from small to large. The other graph plots the change in the forgetting rate caused by the ascending ordered accuracy value corresponding to the top graph. According to the forgetting rate in these figures, we find that no matter what algorithm or strategy is used, the model is bound to have the forgetting issue after incremental learning.

\begin{figure}[h]
\centerline{\includegraphics[width=21pc]{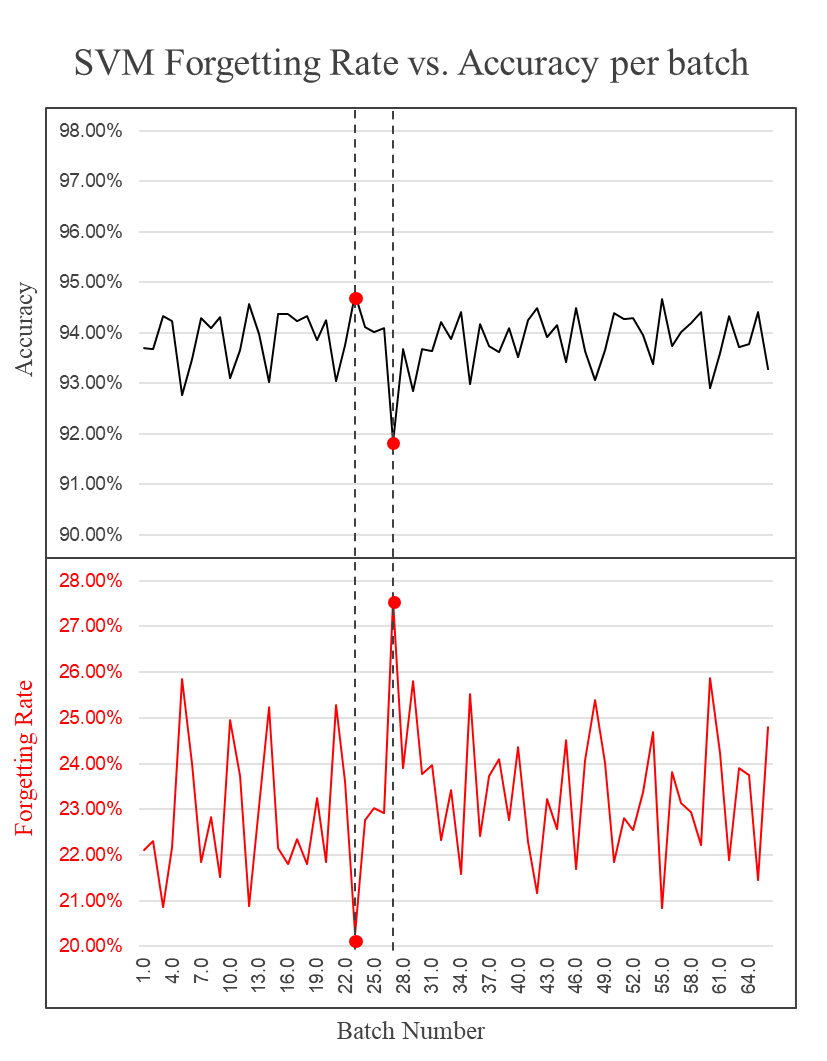}}
\caption{SVM model forgetting rate \& accuracy vs. batch number}
\end{figure}

\begin{figure}[h]
\centerline{\includegraphics[width=21pc]{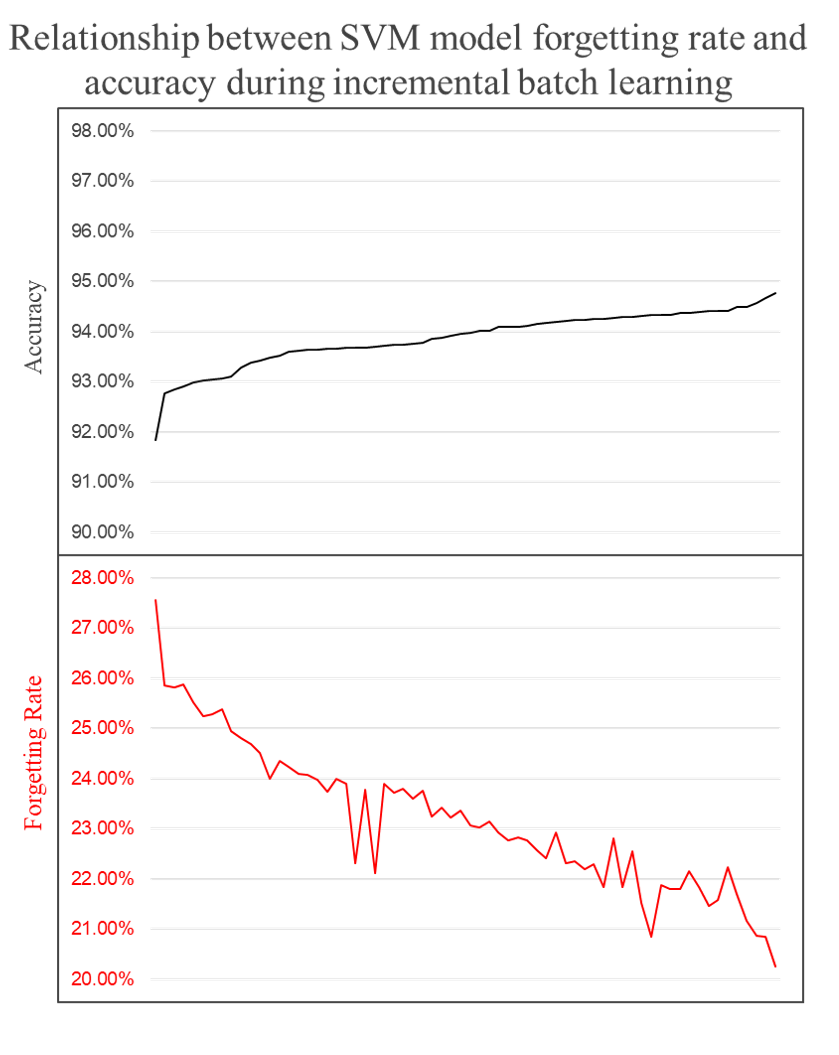}}
\caption{SVM model forgetting rate vs. accuracy}
\end{figure}

\begin{figure}[h]
\centerline{\includegraphics[width=21pc]{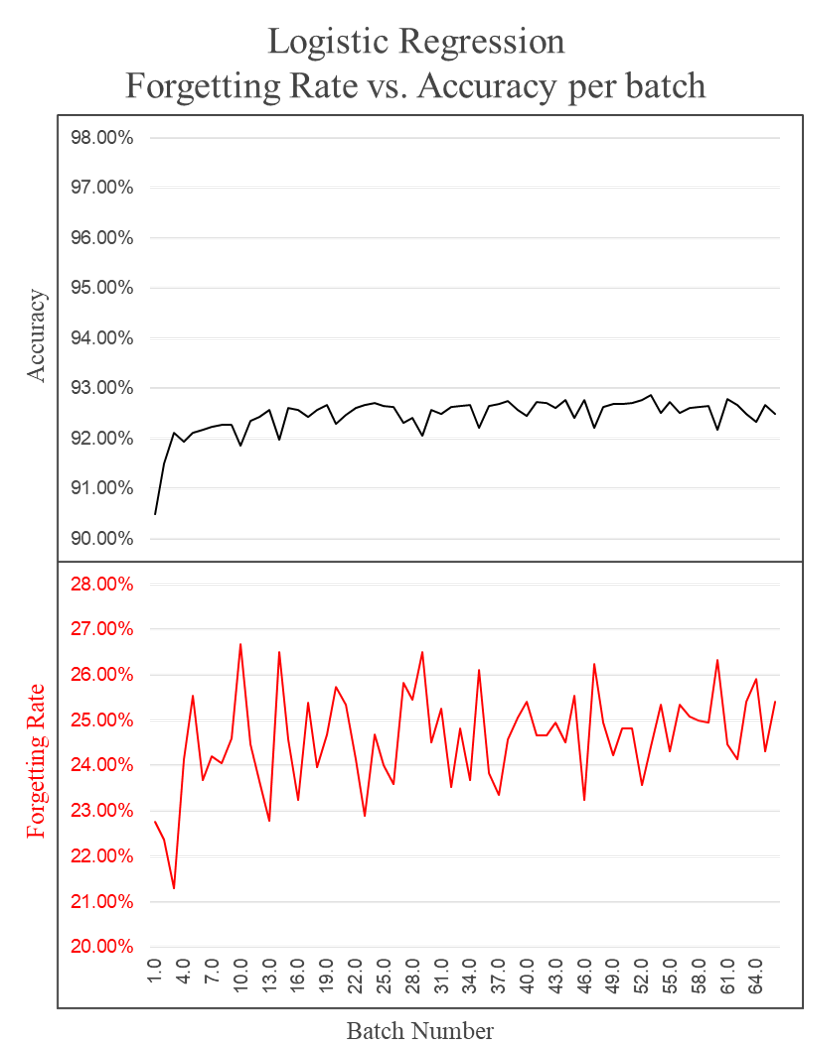}}
\caption{logistic regression model forgetting rate \& accuracy vs. batch number}
\end{figure}

\begin{figure}[h]
\centerline{\includegraphics[width=21pc]{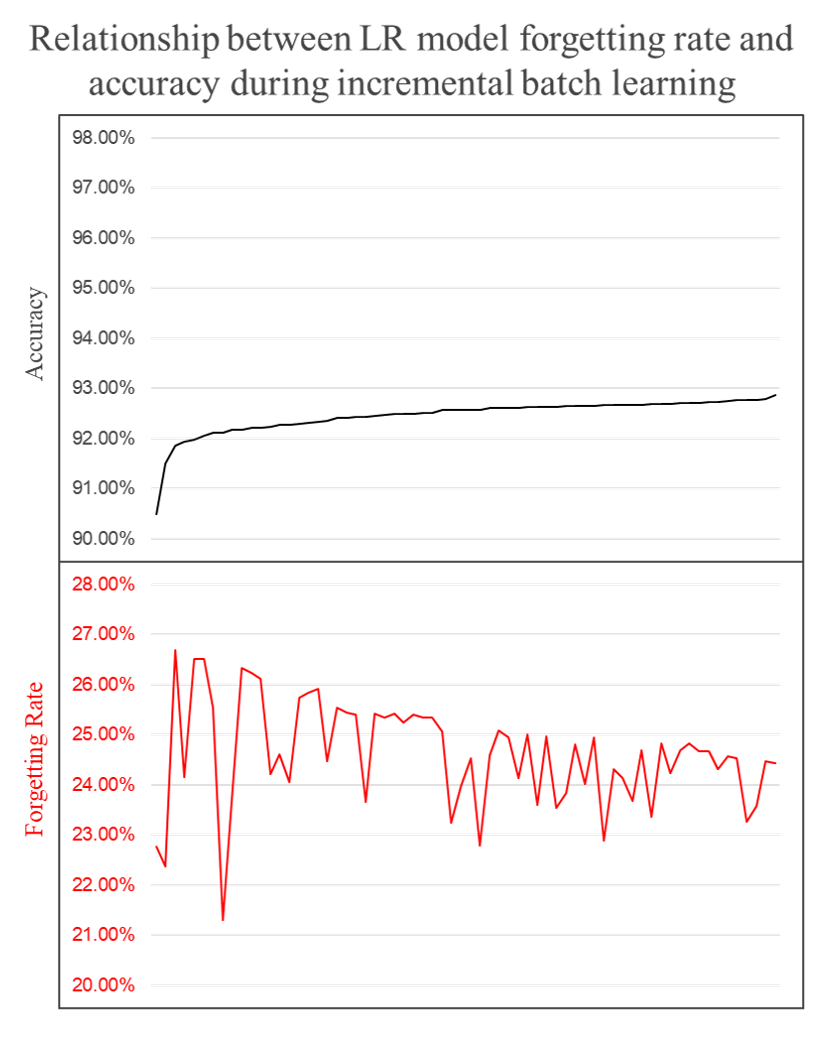}}
\caption{logistic regression model forgetting rate vs. accuracy}
\end{figure}

\begin{figure}[h]
\centerline{\includegraphics[width=21pc]{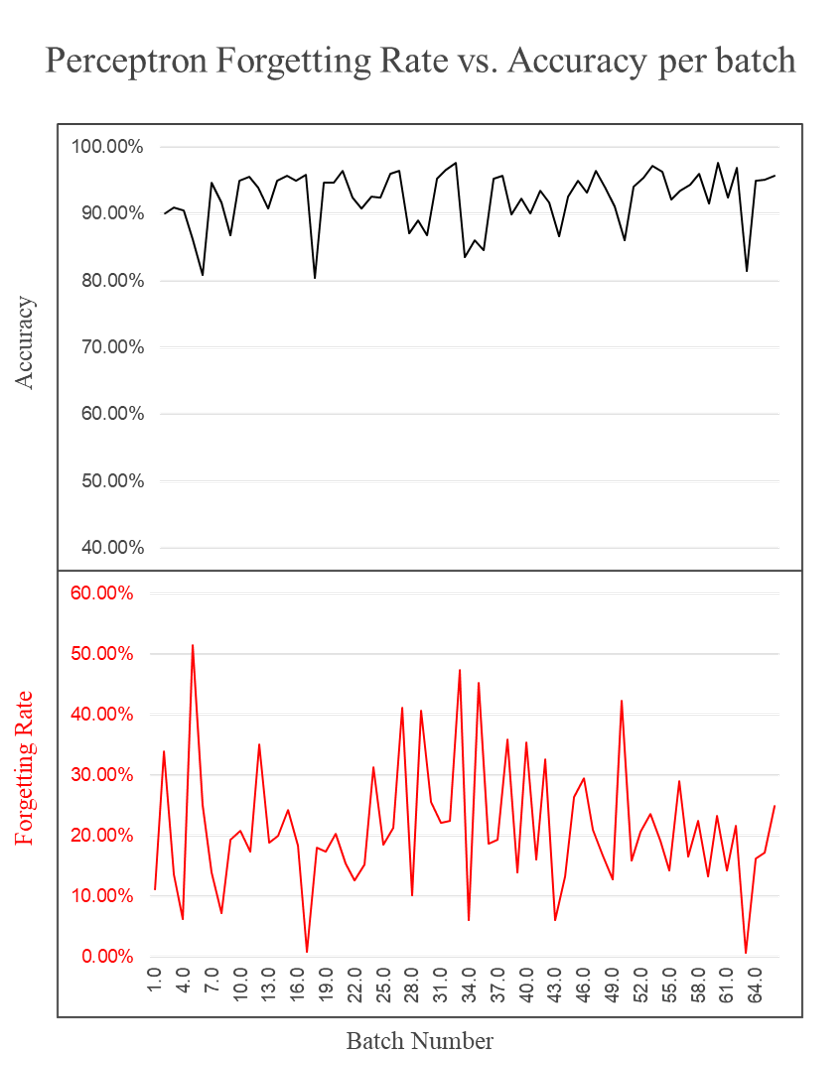}}
\caption{Perceptron model forgetting rate \& accuracy vs. batch number}
\end{figure}

\begin{figure}[h]
\centerline{\includegraphics[width=21pc]{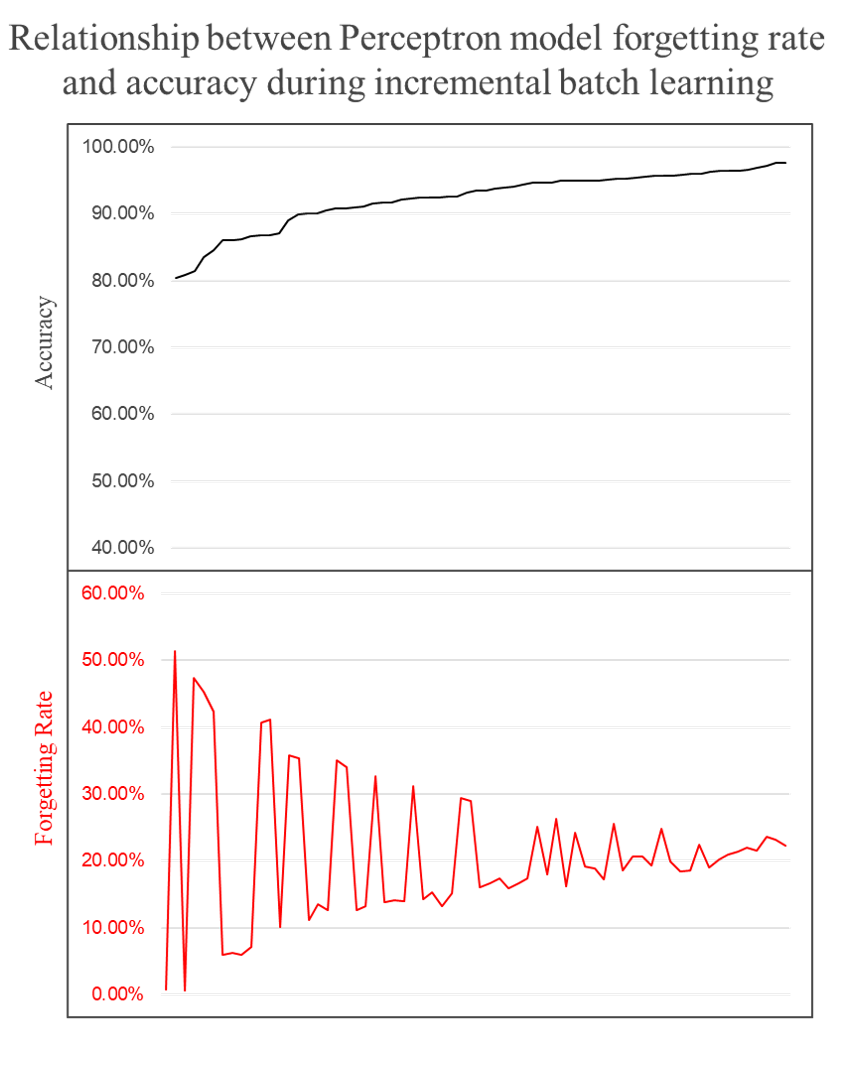}}
\caption{Perceptron model forgetting rate vs. accuracy}
\end{figure}

According to Fig 4, 6, and 8, We can observe that after incremental learning, the accuracy performance of all three models facing new incoming data has improved. However, different models have different relationships between forgetting rate and accuracy. Based on Fig 4 and 5, We can observe that the SVM model achieves a low forgetting rate in the batch number where high accuracy can be achieved. For example, in batch number 27, SVM model achieves 91.85\% accuracy with 27.56\% forgetting rate. However, in batch number 23, SVM model achieves 94.76\% accuracy with 20.26\% forgetting rate. This may indicates that the less the model forgets in the process of incremental training, the more the model can improve the detection accuracy of new incoming data. Compare Fig 6 \& 7 with Fig 4\& 5, the Logistic Regression model performed similarly to the SVM, but its forgetting rate fluctuated less (21\% to 27\%) over 66 batches than the SVM model (20\% to 28\%). 

According to Fig 8 and Fig 9, the changes of forgetting rate and accuracy in the Perceptron model are different from those in the above two models. The fluctuation of the forgetting rate in the perceptron model (0\% to 55\%) is much greater than that of SVM and logistic regression models. However, the Fig 9 plotted graph of ascending accuracy indicates that the higher the accuracy of the model, the smaller the fluctuation it produces. Finally, it converges to around 21\%. Furthermore, by comparing different horizontal coordinate standards in two figures of the same model, we can observe that the accuracy and forgetting rate of each model do not increase with the increase of batch number. This is due to a combination of the batch size, the algorithm, and the dataset we chose. Incremental batch learning with early stopping can be viewed as addressing catastrophic forgetting issues by considering model performance in each batch learning. Certain rules can be set manually to early stop the online batch learning for a trade-off between forgetting rate and other performance evaluations.

In addition to the above early stopping strategy, Learning without forgetting ~\cite{article66} (LwF) was further applied to address forgetting issues based on the knowledge distillation concept. Knowledge distillation concept is to make the knowledge of a teacher model can be transferred to a student model. LwF is a strategy to cope with the catastrophic forgetting problem by utilizing knowledge distillation concept to teach the new model not to forget. It is mainly applied in neural networks and deep learning. Therefore, in addition to the three algorithms used in experiment 1, a deep neural network model, DNN, constructed by 2 dense layers with LwF algorithm design was also tested in experiment 2B.

The core idea of LwF is to try to maintain the learning outcomes of the original task when learning a new task. In the traditional neural network training process, when the network is trained for a new task, the network parameters change to adapt to the new task. Thus, it often leads to a decrease in the learning of the original task, which is catastrophic forgetting. When learning a new task or new data, the model not only optimizes for the loss of the new task or data, but also tries to minimize the change in the prediction result of the original task or data. For instance, LwF tries to keep the prediction result of the model on the original task or data unchanged. Thus, LwF is able to learn the new task or data while preserving the learning effect on the old task or data, so as to effectively avoid or mitigate the problem of catastrophic forgetting. The experiment results are recorded in Table VIII.

\begin{table}[]
\caption{Experiment 2B Result: DNN model with and without LwF under incremental learning}

\begin{tabular}{|l|c|c|}
\hline
\multicolumn{1}{|c|}{DNN Model}              & \begin{tabular}[c]{@{}c@{}}Accuracy \\ without\\ LwF\end{tabular} & \begin{tabular}[c]{@{}c@{}}Accuracy\\ with\\ LwF\end{tabular} \\ \hline
Offline testset before Incremental learning  & 97.21\%                                                           & 97.15\%                                                       \\ \hline
Offline testset after Incremental learning   & 78.35\%                                                           & 82.16\%                                                       \\ \hline
Incoming testset before Incremental learning & 56.21\%                                                           & 56.35\%                                                       \\ \hline
Incoming testset after Incremental learning  & 99.65\%                                                           & 95.05\%                                                       \\ \hline
Forgetting Rate                              & 19.40\%                                                           & 15.43\%                                                       \\ \hline
Offline training time cost                   & 17.15s                                                            & 16.26s                                                        \\ \hline
Incremental learning time cost               & 57.29s                                                            & 52.10s                                                        \\ \hline
\end{tabular}
\end{table}

We can find that the forgetting rate of DNN decreased from 19.4\% to 15.43\% after applying LwF algorithm. However, the accuracy for new incoming data drops from 99.65\% to 95.05\%. This is the trade-off between forgetting rate and accuracy by applying LwF. Distillation Weight ($\lambda$) is an important parameter that is used in the loss function to control the balance between the importance of the new incoming data and old task data. A higher $\lambda$ places more emphasis on remembering the old task data. In this experiment, we set the same distillation weight to loss function values of the updating model and the original offline model. The distillation weight of LwF algorithm can be further adjusted to achieve different trade-offs between forgetting rate and accuracy according to the importance of different evaluation measures.

\section{\bf Conclusion}

The high data rates, coverage, and extremely low latency offered by 5G bring new challenges for network traffic detection while improving network services. The contribution of this paper is to first provide a comprehensive analysis of 5G technologies as well as 5G security. This is followed by a summary and review of SOTA traffic detection techniques and datasets. Then we propose three major issues that need to be addressed in conjunction with 5G and malicious traffic detection: high-performance processing of detection models, data security and privacy, and highly robust detection capabilities. Finally, we propose the use of incremental learning techniques to address the above three issues to some extent and analyze the benefits of applying incremental learning techniques to traffic detection. The experimental results demonstrate that incremental learning can achieve high robustness in detection that cannot be achieved by traditional training methods while achieving high performance in processing, e.g., the incrementally learned SVM model can achieve higher detection accuracy than the same SVM model with traditional offline learning in the face of new emerging data. Finally, we analytically discuss the catastrophic forgetting problem brought by incremental learning itself. The trade-off between the forgetting rate of old data and the robustness of new emerging data in Experiment 2 is also discussed. In the future, we will further investigate incremental learning-based malicious traffic detection in the 5G environment to further improve the performance of the model. More different machine learning and deep learning algorithms and incremental learning strategies will be investigated in depth.

\end{document}